\documentclass[12pt,preprint]{aastex}

\begin{document}

\title{SS Ari: a shallow-contact close binary system}

\author{Liu L.\altaffilmark{1,2,3}, Qian S.-B.\altaffilmark{1,2,3}, He J.-J.\altaffilmark{1,2,3},  Zhang J.\altaffilmark{1,2,3}
 and Li L.-J.\altaffilmark{1,2,3}}

\altaffiltext{1}{National Astronomical Observatories/Yunnan
Observatory, Chinese Academy of Sciences, P.O. Box 110, 650011
Kunming, P.R. China (e-mail: creator\_ll.student@sina.com;
LiuL@ynao.ac.cn)}

\altaffiltext{2}{United Laboratory of Optical Astronomy, Chinese
Academy of Sciences (ULOAC), 100012 Beijing, P. R. China}

\altaffiltext{3}{Graduate School of the CAS, Beijing, P.R. China}

\begin{abstract}
Two CCD epochs of light minimum and a complete R light curve of SS
Ari are presented. The light curve obtained in 2007 was analyzed
with the 2003 version of the W-D code. It is shown that SS Ari is a
shallow contact binary system with a mass ratio $q=3.25$ and a
degree of contact factor $f=9.4\%(\pm0.8\%)$. A period investigation
based on all available data shows that there may exist two distinct
solutions about the assumed third body. One, assuming eccentric
orbit of the third body and constant orbital period of the eclipsing
pair results in a massive third body with $M_3=1.73M_{\odot}$ and
$P_3=87.0$yr. On the contrary, assuming continuous period changes of
the eclipsing pair the orbital period of tertiary is 37.75yr and its
mass is about $0.278M_{\odot}$. Both of the cases suggest the
presence of an unseen third component in the system.

\end{abstract}

\section{Introduction}

SS Ari, which was discovered as a variable star by Hoffmeister
(1934) and then was recognized as a W Ursae Majoris type eclipsing
binary system, has been paid more and more attention to due to its
erratic period variations (Demircan \& Selam 1993) since 1993. Many
high accurate CCD times of minimum of it has been obtained in recent
years. Five years have past since the last researches have been done
by Kim et al. (2003). It is high time that we did some new studies
to compare with the results found out by our predecessors.

Many researchers have presented its photoelectric light, such as
Zhukov (1975), Kaluzny \& Pojmanski (1984a, b), Rainger et al.
(1992), Liu et al. (1993) and Kim et al. (2003). At the same time,
spectroscopic observations for this system have been reported by Lu
(1991), Rainger et al. (1992) and Kim et al. (2003). Furthermore,
the period variabilities were studied by Braune (1970), Kaluzny \&
Pojmanski (1984a, b), Kurpinska-Winiarska \& Zakrzewski (1990),
Rainger et al. (1992), Demircan \& Selam (1993), Kim et al. (1997),
Kim et al. (2003). And so on.

Up to now, it is agreed that SS Ari is a sun-like W-type W UMa
system with a spectral type in the interval F8V-G2V, mass parameters
$M_1=0.4M_{\odot}$, $M_2=1.3M_{\odot}$. As Kim et al. (2003) pointed
out, there are still many discordant matters among the previous
investigations: diverse interpretations of the period activity,
apparent intrinsic variability of the light curve, and disparate
radial velocity curves.  New times of minimum can validate the
previous results about its period changes; new light curves can hope
to make a contribution to the first two confusions. Those are the
motivation of our current work.

\section{New observations}

SS Ari was observed on two nights (December 13, 14, 2007) with the
PI512 TKB CCD photometric system attached to the 60cm reflecting
telescope at the Yunnan Observatory in China. The R color system
used is close to the standard UBVRI system. The effective field of
view of the photometric system is $7\times7$ arcmin at the
Cassegrain focus. The integration time for each image is 15\,s. The
comparison star is BD$+23^{\circ}277$. PHOT (measure magnitudes for
a list of stars) of the aperture photometry package of IRAF was used
to reduce the observed images. Through the observation we obtained a
complete R light. By calculating the phase of the observations with
Equation 2, the light curves are plotted (Figure 1) and the original
data in the R band are listed in Table 1. In this figure, it is
shown that the data are high quality and the light variation is
typical of EW type. Since the lights around the minimum are
symmetric, a parabolic fitting was used to determine the times of
minimum light by the least square method. In all, our two new epochs
of light minimum were obtained and listed in the last line of Table
2.

\section{Orbital period variations}
The orbital period of the W-type contact binary, SS Ari, has been
studied by several authors. Braune (1970) first noted that the
period of SS Ari is variable. And this conclusion was confirmed
later by Kaluzny \& Pojmanski (1984b). They deemed that the period
can be described by two possibilities, (i) continuous sine-like
variation, (ii) change formed by several jumps. Kurpinska-Winiarska
\& Zakrzewski (1990) questioned the sine-like shape of the O-C curve
of SS Ari, since the sine-like variation mainly results from the
positions of the visual and photographic timings published before
1965. They forced to present the O-C observations by a parabolic
fitting, claiming that the sine-like fitting was also acceptable if
the O-C values of the timings published before 1965 were shifted by
half a period. Later, the period changes of SS Ari were studied in
detail by Demircan \& Selam (1993); their results showed that the
O-C diagram of the system varies as a cyclic oscillation of which
the amplitude is $0.^{d}0398$ and the period is 44.8\,yr. This
phenomena was explained as light-time effect, which is caused by a
hypothetical third body with $M_{3}\sim1.0M_{\odot}$. Since not any
sign of the third body was seen in the spectrum of this binary,
Lu(1991) pointed out that the hypothetical third body is too faint,
most probably a binary just as in the case of XY Leo (Barden 1987),
or a white dwarf. The latest researches were given by Kim et al.
(2003). They postulated three possible reasons for the period
variation. First, it was caused by cyclical magnetic activity or by
a light-time effect due to a gravitationally bound third star.
Second, it may be a secular period change because of mass transfer.
Third, the real variations were more complicated. They concluded
that there exist not only a third body
($m_3=1.96M_{\odot},P_3=39.7\rm yr$), but also a fourth body
($m_4=2.38M_{\odot},P_4=88.2\rm yr$).

\subsection{Model I: A constant orbit superimposed a eccentric orbit fitting }
Demircan \& Selam gave their liner ephemeris:
\begin{equation}
MinI=2444469.4790+0^{d}.40599144\times{E}
\end{equation}\noindent
After that investigation, some photoelectric and CCD times of light
minimum have been published. All its available times of light
minimum have been collected to look for the real period changes,
including visual, photographic, photoelectric and CCD observations.
Those were completely listed in Table 2. Based on all collected
eclipse times, setting visual and photographic data as weight 1,
meanwhile setting photoelectric and CCD data as weight 8, a new
linear ephemeris was obtained:
\begin{equation}
MinI=2444469.4879(\pm0.0030)+0^{d}.40598629(\pm0.00000026)\times{E}.
\end{equation}
\noindent The $(O-C)$ values with respect to the linear ephemeris
are listed in the fifth column of Table 2. The corresponding $(O-C)$
diagram is displayed in Figure 2. It is no doubt that the general
$(O-C)$ trend of SS Ari, shown in Figure 2, is a cyclic variation.
Consider the relationship of the two periods derived by Kim et al.
(2003), an eccentric orbit and constant period fitting was used. The
result is:
\begin{eqnarray}
MinI&=&2444469.4169+0.40598629\times{E} \nonumber\\
&&+0.0859\cos0^{\circ}.0046E+0.0595\sin0^{\circ}.0046E\nonumber\\
&&+0.0023\cos0^{\circ}.0092E-0.0205\sin0^{\circ}.0092E.
\end{eqnarray}
 \noindent The corresponding O-C curve is shown in
Figure 2, where solid cycles refer to the photoelectric or CCD
primary minima and open ones to the photoelectric or CCD secondary
minima. Solid triangles denote visual or photographic primary
minima, open triangles denote visual or photographic secondary
minima; meanwhile, solid line represents an eccentric ephemeris
variation. The corresponding residuals were drawn in Figure 3. From
it we can see all photoelectric, CCD and most of visual,
photographic timings' residuals are in a horizontal line, except
some scatters. It accounts for that the line fits all photoelectric
and CCD data very well, although the visual and photographic data
show large scatters in the E interval 0-10000. By aid of relations
(Kopal 1978),
\begin{equation}
\omega=360^{\circ}P_{e}/T,
\end{equation}
\begin{equation}
e^{\prime}=2\sqrt{\frac{a_2^2+b_2^2}{a_1^2+b_1^2}},
\end{equation}
\begin{equation}
w^{\prime}=arctan
\frac{(b_1^2-a_1^2)b_2+2a_1b_1a_2}{(a_1^2-b_1^2)a_2+2a_1b_1b_2},
\end{equation}
\noindent where $P_{e}$ is the ephemeris period ($0.^{d}40598629$)
and $a_1, a_2, b_1, b_2$ are the corresponding fitting coefficients
($a_1=0.0859, b_1=0.0595, a_2=0.0023, b_2=-0.0205$), the period of
the orbital oscillation was determined to be T=87.0yr with an
eccentricity $e^{\prime}=0.3948$. These may suggest that there
exists a small-amplitude oscillation in the period changes, which
can be explained by the presence of an unseen third body in the
system. This unseen third body was first proposed by Demircan \&
Selam (1993) and later confirmed by Kim et al. (2003).

\subsection{Model II: A quadratic change superimposed a cycle change fitting }
However, since very small amount of the early data (before 1960),
the photographic and visual times of light minimum showed a large
scatter (up to 0.06 days). Those data not only do not contribute to
form the general O-C trend of the binary system but also tend to
mislead regarding the real period changes. Considered this, the key
visual data in early years may be fallible, the simulation by
eccentric orbit and constant period model may be incredible. The
same situation was met during the period analysis of AM Leo (Qian et
al. 2005) and HL Aur (Qian et al. 2006), AP Leo (Qian et al. 2007).
This situation suggests that some of the period variations pointed
out by former investigators are hardly reliable. Therefore, only
using those photoelectric and CCD times of minimum light for the
present period study of SS Ari is necessary. The new linear
ephemeris purely derived by photographic and CCD data is:
\begin{equation}
MinI=2444469.4789(\pm0.0071)+0^{d}.40599166(\pm0.00000053)\times{E}.
\end{equation}
\noindent Based on these high accurate data, we give a quadratic
fitting:
\begin{eqnarray}
MinI&=&2444469.4988(\pm0.0004)+0.40598838(\pm0.00000002)\times{E} \nonumber\\
&&-1.72(\pm0.01)\times{10^{-10}}E^2.
\end{eqnarray}
\noindent That fitting is shown in Figure 4, the dash line.
Actually, only a quadratic simulation can not fit the data well.
Having bethought of the doubtless cyclic oscillation in O-C diagram,
we used a circular orbit and continuous period change form to fit
these data, the result is:
\begin{eqnarray}
MinI&=&2444469.4999(\pm0.0016)+0.40598942(\pm0.00000020)\times{E} \nonumber\\
&&-2.24(\pm0.10)\times{10^{-10}}E^2+0.0112(\pm0.0012)\sin(0^{\circ}.0106E-31^{\circ}.5).
\end{eqnarray}
\noindent The resultant O-C diagrams are also shown in Figure 4 and
5, as the solid line represents. That fitting gives a long-term
decrease ($dP/dt=-4.03\times{10^{-7}}$\,days/year) with a clear
period oscillation ($A_3=0.^{d}0112$, $T_3=37.75\,years$). The
residuals were plotted in Figure 6. Comparing with the Figure 3, the
residuals in this graph is purely good.

\section{Photometric solutions}
Because of its unusual period changes, spectroscopic observation
have been done by Lu (1991) and Rainger et al. (1992). Ten years
later, Kim et al. (2003) gave a comprehensive investigation,
including the period changes, the photometric and spectroscopic
parameters. According to their studies, the spectroscopic mass ratio
is more likely $q=3.25$ which we adopted in our photometric
solution. Of course, as a standard process, to check this value, a
q-search method with the 2003 version of the W-D program (Wilson \&
Devinney, 1971, Wilson, 1990, 1994, Wilson \& Van Hamme, 2003) was
used (Figure 7). We fixed q to 0.3, 0.4, 0.5, and so on, as figure 7
shows. It can be seen that the best value is between $q=3.1$ and
$q=3.7$, which is acceptable in the range of the errors.

During the solution, the temperature of star 1 (star eclipsed at
primary light minimum) was fixed at $T_1=5860$K, the same value used
by Kim et al. (2003). The bolometric albedo $A_1=A_2=0.5$ (Rucinski
1969) and the values of the gravity-darkening coefficient
$g_1=g_2=0.32$ (Lucy 1967) were used, which correspond to the common
convective envelope of both components. According to Claret \&
Gimenez (1990), a limb-darkening coefficient of 0.491 in R was used.
We adjusted the orbital inclination $i$; the mean temperature of
star 2, $T_2$; the monochromatic luminosity of star 1, $L_{1R}$ and
the dimensionless potential of star 1 ($\Omega_1=\Omega_2$, mode 3
for contact configuration). A small O'Connell effect of the system
can not be ignored. As SS Ari's spectral type is in the interval
F8V-G2V, a sun-like star, it seemed that the probability of
appearance of starspots on the surface of the star is high. In fact,
as many researchers have done before (e.g., Liu et al. 1993, Kim et
al. 2003), which were an agreement now, we add a spot on the star 2
that was subsequently proved as the cooler more massive primary
component. The photometric solutions are listed in Table 3 and the
theoretical light curves computed with those photometric elements
are plotted in Figure 8. In order to get a image of the binary and
its surface spots in our mind, the geometrical structure of SS Ari
is displayed in Figure 9. For comparing, the results of previous
studies are listed in Table 4.

\section{Discussions and conclusions}

The orbital period was revised as 0.40599166 days by using 115
photoelectric and CCD timings of SS Ari listed in Table 2. This
system is a shallow contact W-Type binary with $q=3.25$, $f=9.4\%$.
These suggest that the system may be a marginal contact binary. The
temperatures of the two components are different (372K), that
phenomena is very common in marginal contact binaries. Marginal
contact binaries, whose fill factor are very small, ($f\leqslant
10\%$) are indicators of the time-scale that a binary will spend to
evolve into the contact stage. For example, II CMa (Liu et al.
2008), V803 Aql (Samec et al. 1993), FG Sct (Bradstreet 1985), RW
PsA (Lucy \& Wilson 1979), XZ Leo (Niarchos et al. 1994), S Ant
(Russo et al. 1982).

\subsection{Model I}
As shown in Figures 2, 3 and 4, both of the primary and the
secondary times of light minimum follow the same general trend of
O-C variation indicating that the weak O-C oscillation can barely be
explained as apsidal motion. The alternate period change of a close
binary containing at least one solar-type component at least can be
interpreted by the mechanism of magnetic activity (e.g., Applegate
1992, Lanza et al. 1998). However, for contact binary stars, we
scarcely know whether this mechanism can work or how it might work;
the common convective envelope covers the truth. We deduce that the
period oscillation may be caused by the light-time effect of a
tertiary component. The trend of O-C variation is very clearly, as
we can see from Figure 2, though the data between E=0 and E=10000
show large scatters. In the previous section, by assuming an
eccentric orbit and constant period model, a theoretical solution of
the orbit for the assumed tertiary star was calculated. By using
this equation:
\begin{equation}
f(m)=\frac{4{\pi}^{2}} {{\it
G}T_3^{2}}\times(a_{12}^{\prime}\sin{i}^{\prime})^{3},
\end{equation}
(where
$a_{12}^{\prime}\sin{i}^{\prime}=A_3/\cos{w}^{\prime}\times{c}/(1\pm
e^{\prime})$; c is the speed of light; $e^{\prime}$ is the
eccentricity of the third body's orbit; $w^{\prime}$ is the
longitude of periastron of the third body's orbit), the mass
function for the tertiary component is computed. Then, with the
following equation:
\begin{equation}
f(m)=\frac{(M_{3}\sin{i^{\prime}})^{3}} {(M_{1}+M_{2}+M_{3})^{2}},
\end{equation}
taking in the physical parameters $M_1=1.3M_{\odot}$,
$M_2=0.4M_{\odot}$ (Kim et al. 2003), the masses and the orbital
radii of the third companion are computed. The values for several
different orbital inclinations ($i^{\prime}$) are shown in Table 5.
If the tertiary companion is coplanar to the eclipsing pair
(i.e.,with the same inclination as the eclipsing binary), its mass
should be $m_3=1.73M_{\odot}$. This mass is big enough to be
detected. However, several spectral observations did not find a
third body any more. Why is so massive star difficult to be seen?
The possibility is that it is a compact star, what consistent in the
results given by Kim et al. (2003). If the compact object really
exists, the most point is how can the third body evolve into a
compact star leaving a main sequence centric binary system? The
interpretation by capture theory is more reasonable than by forming
at same time and evolving together.

\subsection{Model II}

However, from Eq.9, we were amazed at the very small mass of the
third body (About $0.278M_{\odot}$, see Table 5). We think that this
model is a much better interpretation of the unseen tertiary. All
doubts mentioned above will be swept away as the unseen component is
a small mass star (e.g., the third body should be a cool dwarf star
which is unseen spectroscopically and contributes extremely low
light to the total system in optical band). This fitting gives its
long-term decrease with a change rate
$dP/dt=-4.03\times{10^{-7}}$\,days/year, which may be due to a
conservative mass transfer from the more massive component to the
less massive one. Then with the accepted absolute parameters, the
well-known equation (Tout \& Hall 1991),
\begin{equation}
\frac{\dot{P}}{P}=3\frac{\dot{M_2}}{M_2}(\frac{M_2}{M_1}-1),
\end{equation}
\noindent the mass transfer rate is estimated to be
$dM_2/dt=-1.91\times{10^{-7}}M_{\odot}/year$ where minus expresses
the more massive component $M_2$ losing mass. The timescale of mass
transfer is $\tau\sim{M_2/\dot{M_2}}\sim6.80\times{10^{6}}$ years.
This number is half of the thermal time scale of the massive
component. On the other hand, both components of SS Ari are
solar-type stars (spectral type in the interval F8V-G2V) but they
rotate about 100 times as fast as the Sun, indicating a high degree
of magnetic activity from the spin-up of the components. The
asymmetry and variation of the light curve are indicators of spot
activity. The long term decrease of the orbital period can be
explained as the results of an enhanced stellar wind and AML. It is
possible that the cyclic period change is caused by magnetic
activity-driven variations in the quadrupole momentum of the
solar-type components (e.g., Applegate 1992, Lanza et al. 1998).

Qian (2001a, b, 2003a) has shown the long-term period variation of
contact binary stars may correlate with the mass of the primary
component ($M_{1}$), the mass ratio of the system ($q$). His
statistic critical mass ratio q is 0.4. When $q>0.4$, the secular
period increases; contrary, $q<0.4$, the secular period decreases.
The secular period decrease of SS Ari is consistent with this
conclusion. To interpret the secular period changes of contact
binary stars, Qian (2001a, b, 2003a) has proposed an evolutionary
scenario of eclipsing binary stars. According to this scenario, the
evolution of a contact binary may be the combination of the thermal
relaxation oscillation (TRO) and the variable angular momentum loss
(AML) via the change of depth of contact.  Systems (e.g., V417 Aql,
see Qian 2003b) with a secular decreasing period are on the
AML-controlled stage, while those (e.g., CE Leo, see Qian 2002)
showing an increasing period are on the TRO-controlled stage. The
long-term period decrease of SS Ari may suggest that it is on the
AML-controlled stage of this evolutionary scheme.

%\bigskip
\acknowledgments{This work is partly supported by Yunnan Natural
Science Foundation Foundation (No.2005A0059M), Chinese Natural
Science Foundation (No.10573032, No.10433030, No.10573013 and
No.10778707), The Ministry of Science and Technology of the
People¡¯s Republic of China through grant 2007CB815406 and The
Chinese Academy of Sciences grant No.O8ZKY11001 and No.O8AXB51001.
New observations of the system were obtained with the 60cm telescope
at Yunnan Observatory. Thanks to the anonymous referee who has given
us very constructive comments and cordial suggestions, which helped
us to improve the paper greatly.}

\begin{table*}
\begin{tiny}
\caption{The original data of SS Ari in R band observed by 60cm
telescope at Yunnan Observatory. Hel. JD 2454400+}
\begin{tabular}{llllllllll llll}
\hline
 Hel.JD&$\Delta{m}$&Hel.JD&$\Delta{m}$&Hel.JD&$\Delta{m}$&Hel.JD&$\Delta{m}$&Hel.JD&$\Delta{m}$&Hel.JD&$\Delta{m}$&Hel.JD&$\Delta{m}$\\
 \hline
47.9884 &   1.158 & 48.0343 &   0.823 & 48.0799 &   0.713 & 48.1287 &   0.798 & 48.1754 &   1.150 & 48.2281 &   0.899 & 48.2861 &   0.674 \\
47.9897 &   1.151 & 48.0348 &   0.824 & 48.0804 &   0.705 & 48.1292 &   0.797 & 48.1760 &   1.157 & 48.2290 &   0.884 & 48.2872 &   0.665 \\
47.9902 &   1.155 & 48.0354 &   0.819 & 48.0810 &   0.704 & 48.1298 &   0.794 & 48.1766 &   1.149 & 48.2295 &   0.874 & 48.2880 &   0.693 \\
47.9913 &   1.146 & 48.0360 &   0.816 & 48.0816 &   0.705 & 48.1303 &   0.795 & 48.1772 &   1.162 & 48.2302 &   0.882 & 48.2888 &   0.690 \\
47.9919 &   1.148 & 48.0366 &   0.809 & 48.0821 &   0.699 & 48.1311 &   0.809 & 48.1777 &   1.162 & 48.2313 &   0.877 & 48.2895 &   0.666 \\
47.9924 &   1.143 & 48.0372 &   0.814 & 48.0828 &   0.689 & 48.1318 &   0.807 & 48.1783 &   1.173 & 48.2320 &   0.856 & 48.2901 &   0.671 \\
47.9929 &   1.141 & 48.0378 &   0.805 & 48.0834 &   0.692 & 48.1323 &   0.809 & 48.1789 &   1.175 & 48.2331 &   0.855 & 48.2907 &   0.668 \\
\end{tabular}\end{tiny}
\end{table*}

\begin{tiny}
\begin{tabular}{llllll llllll}
\hline
Hel.JD&$\Delta{m}$&Hel.JD&$\Delta{m}$&Hel.JD&$\Delta{m}$&Hel.JD&$\Delta{m}$&Hel.JD&$\Delta{m}$&Hel.JD&$\Delta{m}$\\
\hline
48.3472 &   0.788 & 49.0971 &   0.690 & 49.1242 &   0.698 & 49.1514 &   0.776 & 49.1788 &   0.955 & 49.2089 &   1.126 \\
48.3477 &   0.728 & 49.0974 &   0.672 & 49.1245 &   0.704 & 49.1517 &   0.778 & 49.1791 &   0.967 & 49.2092 &   1.134 \\
48.3484 &   0.873 & 49.0978 &   0.681 & 49.1248 &   0.702 & 49.1521 &   0.795 & 49.1794 &   0.958 & 49.2095 &   1.132 \\
48.3491 &   0.806 & 49.0982 &   0.686 & 49.1251 &   0.711 & 49.1524 &   0.773 & 49.1797 &   0.982 & 49.2098 &   1.122 \\
49.0706 &   0.732 & 49.0987 &   0.688 & 49.1255 &   0.717 & 49.1527 &   0.797 & 49.1800 &   0.967 & 49.2102 &   1.119 \\
49.0710 &   0.734 & 49.0991 &   0.673 & 49.1258 &   0.704 & 49.1531 &   0.781 & 49.1804 &   0.969 & 49.2105 &   1.123 \\
49.0714 &   0.736 & 49.0994 &   0.683 & 49.1262 &   0.706 & 49.1534 &   0.778 & 49.1807 &   0.971 & 49.2108 &   1.144 \\
\hline
\end{tabular}
\end{tiny}

%\begin{table}
%\begin{footnotesize}
\begin{deluxetable}{lllrrrl}
\tablewidth{0pc} \tabletypesize{\footnotesize} \tablecaption{Times
of light minimum of SS Ari.}
%\begin{tabular}{llllrrrl}
%\hline
\tablehead{ \colhead{Hel.JD}  & \colhead{Type}  & \colhead{Method} &
\colhead{E}  & \colhead{$(O-C)$ Eq.2} & \colhead{$(O-C)$ Eq.7} &
\colhead{Reference} }
%\hline
\startdata
2430948.3290  &  I   &  vis  & -33304     & -0.1915   & -       &  Odynskaya (1949)                                  \\
2432455.3470  &  I   &  vis  & -29592     & -0.1946   & -       &  Kramer (1948)                                     \\
2432455.5520  &  II  &  vis  & -29591.5   & -0.1926   & -       &  Kramer (1948)                                     \\
2432786.2290  &  I   &  vis  & -28777     & -0.1914   & -       &  Kramer (1948)                                     \\
2432786.4300  &  II  &  vis  & -28776.5   & -0.1934   & -       &  Kramer (1948)                                     \\
2435721.5450  &  I   &  pg   & -21547     & -0.1563   & -       &  Huth (1964)                                       \\
2436075.5810  &  I   &  pg   & -20675     & -0.1404   & -       &  Huth (1964)                                       \\
2439028.3870  &  I   &  vis  & -13402     & -0.0726   & -       &  Braune (1970)                                     \\
\enddata
\end{deluxetable}

\begin{table}

\caption{Photometric solutions for SS Ari.}
\begin{tabular}{lcl}
\hline
Parameters              &  Photometric elements  &  errors\\

                        &  spotted solutions   &        \\
\hline
$g_1=g_2$               &     0.32               & assumed\\
$A_1=A_2$               &     0.50               & assumed\\
$x_{1R}=x_{2R}$         &     0.491              & assumed\\
$T_1$                   &     5860K              & assumed\\
$q$                     &     3.250              & assumed\\
$\Omega_{in}$           &     6.9452             & --     \\
$\Omega_{out}$          &     6.3227             & --     \\
$T_2$                   &     5488K              & $\pm12$K \\
$i$                     &     77.306             & $\pm0.205$ \\
$L_1/(L_1+L_2)(R)$      &     0.2814             & $\pm0.0114$ \\
$\Omega_1=\Omega_2$     &     6.8869             & $\pm0.0093$\\
$r_1(pole)$             &     0.2669             & $\pm0.0006$\\
$r_1(side)$             &     0.2785             & $\pm0.0007$\\
$r_1(back)$             &     0.3142             & $\pm0.0013$\\
$r_2(pole)$             &     0.4576             & $\pm0.0006$\\
$r_2(side)$             &     0.4925             & $\pm0.0008$\\
$r_2(back)$             &     0.5190             & $\pm0.0071$\\
$f$                     &     $9.4\,\%$         & $\pm0.8\,\%$\\
$\theta$ ($^\circ$)     &     29.21                            \\
$\psi$($^\circ$)        &     284.50                          \\
$\Omega$($^\circ$)      &     12.755                           \\
$T_s/T_*$               &     0.8333                           \\
$\sum{\omega_i(O-C)_i^2}$ &   0.0029038                         \\
\hline
\end{tabular}
\end{table}

\begin{table}
\caption{The comparison of photometric solutions for SS Ari.}
\begin{tabular}{lccc}
\hline
Parameters              &  Photometric elements &  Photometric elements  &  Photometric elements\\
                        &  unspotted solutions  &  unspotted solutions   &  spotted solutions   \\
                        &  Lu 1991              &  Liu et al. 1993       &  Kim et al. 2003       \\
\hline
$g_1=g_2$               &     0.32              &     0.32               &     0.32             \\
$A_1=A_2$               &     0.50              &     0.50               &     0.50             \\
$x_{1B}=x_{2B}$         &     0.74              &     0.700              &     --               \\
$x_{1V}=x_{2V}$         &     0.61              &     0.560              &     --               \\
$T_1$                   &     5950K             &     6050K              &     6144K            \\
\hline
\end{tabular}
\end{table}

\begin{table*}
\begin{minipage}{12cm}
\caption{The masses and orbital radii of the assumed third body in
SS Ari. The calculated value about $M_3$ and $a_3$ are effective
numbers. }
\begin{tabular}{lcccl}
      \hline
Parameters && Third body & &Units\\
\hline
&eccentric adding cycle && quadratic adding cycle\\
$A_3$      & $0.1045$ &$0.1045$ &$0.0112$ & day\\
$T_3$      & $87.0$ &$87.0$&$37.75$& yr\\
$e^{\prime}$ & $0.3948$& $0.3948$& $0 $\\
$w^{\prime}$& $-26^{\circ}.983$& $-26^{\circ}.983$& $0^{\circ}$\\
$a_{12}{\rm sin}i^{\prime}$ & $14.5566$ &$33.5485$ &$1.9392$&{\rm AU}\\
$f(m)$   & $0.4075$  & $4.9886$  &$5.1172\times10^{-3}$ & $M_{\odot}$\\
$m_{3}(i^{\prime}=90^{\circ})$  & $1.66   $  &  $7.50    $ &$0.271 $&$M_{\odot}$ \\
$m_{3}(i^{\prime}=77.306^{\circ})$  & $1.73   $  &  $8.62    $ &$0.278 $&$M_{\odot}$ \\
$a_{3}(i^{\prime}=90^{\circ})$  & $26.2  $ &  $15.9  $ &$10.1$ &AU \\
$a_{3}(i^{\prime}=77.306^{\circ})$  & $26.5  $ &  $16.5  $ &$10.1$ &AU \\
        \hline
\end{tabular}
\end{minipage}
\end{table*}

\begin{figure}
\begin{center}
\includegraphics[angle=0,scale=0.8 ]{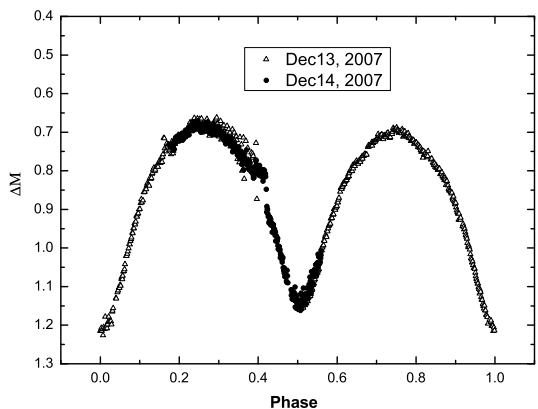}
\caption{CCD photometric light curves in R band of SS Ari obtained
by 60cm reflecting telescope on December 13, 14, 2007.}
\end{center}
\end{figure}

\begin{figure}
\begin{center}
\includegraphics[angle=0,scale=0.8 ]{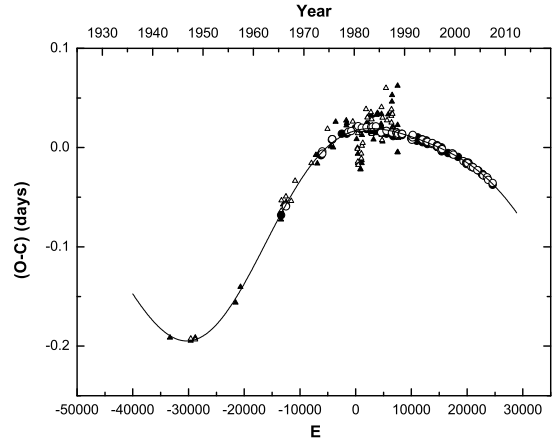}
\caption{$(O-C)$ diagram of SS Ari formed by all available
observations. The $(O-C)$ values were computed by using linear
ephemeris (Eq.2). Solid cycles are referred to the photoelectric and
CCD primary minima while open ones to the photoelectric and CCD
secondary minima. Solid triangles denote visual and photographic
primary minima, open triangles denote visual and photographic
secondary minima, these data may not precise. Solid line represents
an eccentric ephemeris variation (Eq.3).}
\end{center}
\end{figure}

\begin{figure}
\begin{center}
\includegraphics[angle=0,scale=.8]{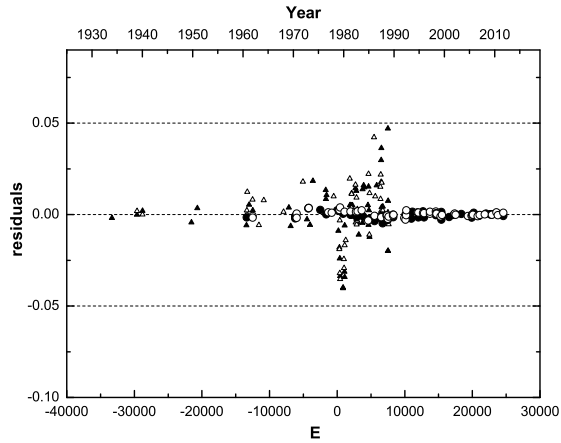}
\caption{The residuals for the eccentric ephemeris (Eq.3). The
symbols are the same as Figure 2.}
\end{center}
\end{figure}

\begin{figure}
\begin{center}
\includegraphics[angle=0,scale=.8]{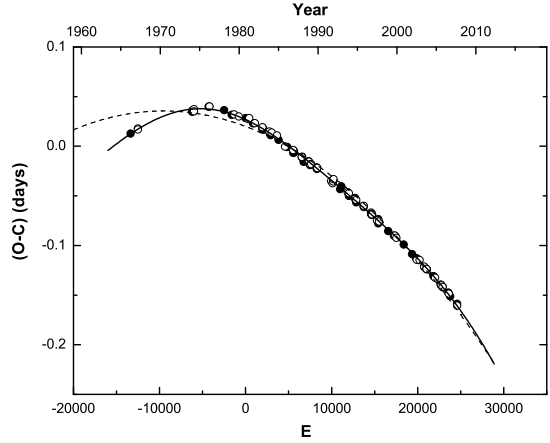}
\caption{Diagram of SS Ari formed by all available photoelectric and
CCD observations. The $(O-C)$ values were computed by using a new
determined linear ephemeris (Eq.7). The symbols are the same as
Figure 2. Solid line represents a quadratic superimposed a cycle
ephemeris variations (Eq.9), meanwhile dash line represent the
quadratic variations (Eq.8).}
\end{center}
\end{figure}

\begin{figure}
\begin{center}
\includegraphics[angle=0,scale=.8]{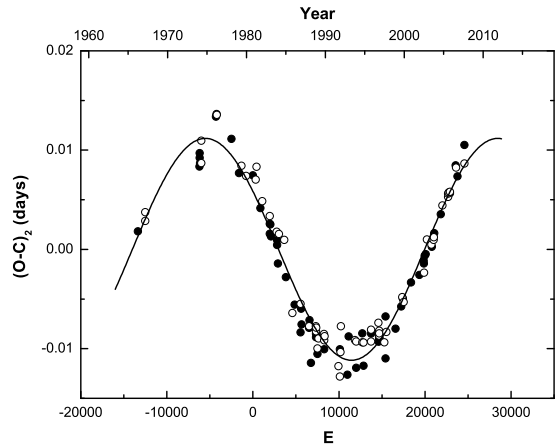}
\caption{$(O-C)_{2}$ values for SS Ari with respect to the cyclic
ephemeris in Eq.(9). The symbols are the same as figure 2. Solid
line refers to the theoretical orbit of an assumed third body.}
\end{center}
\end{figure}

\begin{figure}
\begin{center}
\includegraphics[angle=0,scale=.8]{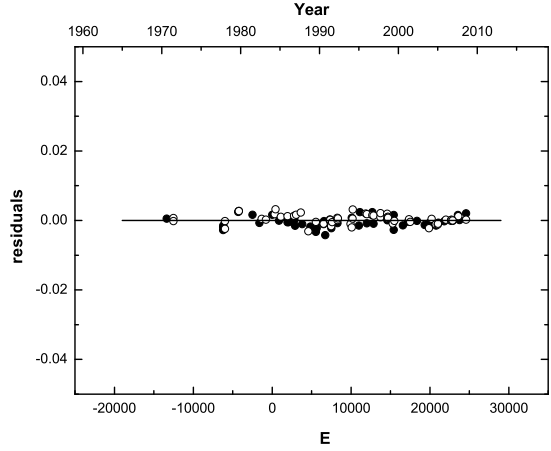}
\caption{The residuals for the quadratic superimposed a cycle
ephemeris and a cyclic variations (Eq.9). The symbols are the same
as Figure 2.}
\end{center}
\end{figure}

\begin{figure}
\begin{center}
\includegraphics[angle=0,scale=.8]{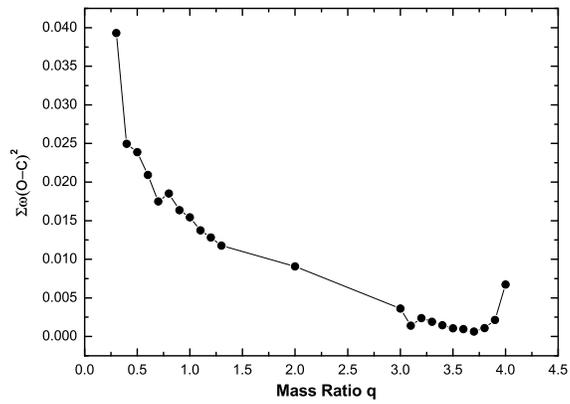}
\caption{The relation between q and $\Sigma$ for SS Ari. It hard to
find a accurate mass ratio from the picture.}
\end{center}
\end{figure}

\begin{figure}
\begin{center}
\includegraphics[angle=0,scale=.8]{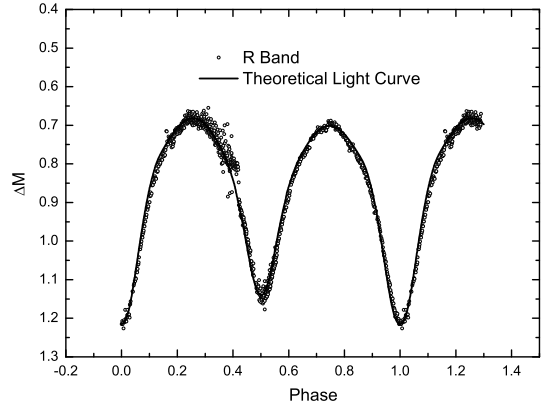}
\caption{Observed (open circles) and theoretical (solid lines) light
curves in R band for SS Ari, with a dark spot in the more massive
component. }
\end{center}
\end{figure}

\begin{figure}
\begin{center}
\includegraphics[angle=0,scale=.8]{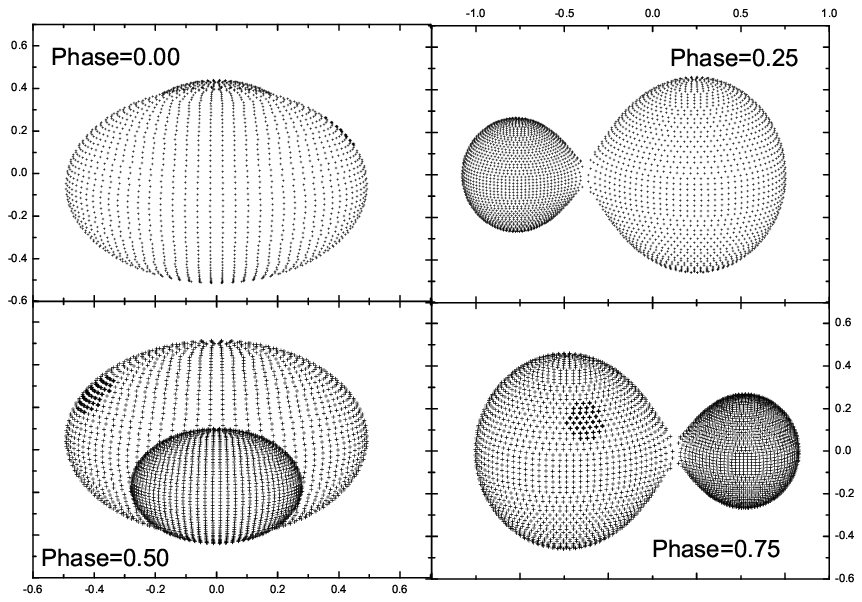}
\caption{Geometrical structure of the shallow contact binary SS Ari
with a dark spot on the more massive component at phase 0.00, 0.25,
0.50 and 0.75.}
\end{center}
\end{figure}


\begin{references}

%
\reference{}Agerer, F., Hubscher, J., 2003, IBVS 5484, 1A
%
\reference{}Applegate, J. H., 1992, ApJ 385, 621
%
\reference{}Barden, S. C., 1987, ApJ 317, 333
%
\reference{}Biro, I. B., Borkovits, T., Hegedus, T., Kiss, Z. T.,
Kovacs, T., Lampens, P., Regaly, Zs., Robertson, C. W., van
Cauteren, P., 2007, IBVS 5760, 1N
%
\reference{}Borkovits, T., Biro, I. B., Hegedus, T., Csizmadia, S.,
Szabados, L., Pal, A., Posztobanyi, K., Konyves, V., Kospal, A.,
Csak, B., Meszaros, S., 2003, IBVS 5434, 1B
%
\reference{} Bradstreet, D. H., 1985, ApJS 58, 413B
%
\reference{}Braune, W., 1970, IBVS 440
%
\reference{}Braune, W., Hubscher, J. \& Mundry, E., 1972, AN 294,
123
%
\reference{}Braune, W. \& Mundry, E., 1973, AN 294, 225
%
\reference{}Braune, W., Hubscher, J. \& Mundry, E., 1977, AN 298,
121
%
\reference{}Braune, W., Hubscher, J. \& Mundry, E., 1979, AN 300,
165
%
\reference{}Braune, W., Hubscher, J. \& Mundry, E., 1981, AN 302, 53
%
\reference{}Braune, W., Hubscher, J. \& Mundry, E., 1983, BAV Bull.
No. 36
%
\reference{}Claret, A., and Gimenez, A., 1990, A\&A 230, 412
%
\reference{}Demircan, O. and Selam S. O., 1993, A\&A 267, 107
%
\reference{}Diethelm, R., 1992, BBSAG Bull. No. 99
%
\reference{}Diethelm, R. \& Blattler, E., 1998, BBSAG Bull. No. 116
%
\reference{}Faulkner, D. R., 1986, PASP 98, 690
%
\reference{}Hanzl, D. R., 1990, IBVS 3423
%
\reference{} Hegedus, T., Biro, I. B., Borkovits, T., and Paragi,
1996, IBVS 4340
%
\reference{}Hoffmeister, C., 1934, Astron. Nachr. 253, 195
%
\reference{}Hubscher, J., 2005, IBVS 5643, 1H
%
\reference{}Hubscher, J. \& Mundry, E., 1984, BAV Bull. No. 38
%
\reference{}Hubscher, J., Lichtenknecker, D. \& Meyer, J., 1986, BAV
Bull. No. 43
%
\reference{}Hubscher, J. \& Lichtenknecker, D., 1988, BAV Bull. No.
50
%
\reference{}Hubscher, J., Agerer, F. \& Wunder, E., 1992, BAV Mitt.
No. 60
%
\reference{}Hubscher, J., Agerer, F. \& Wunder, E., 1993, BAV Mitt.
No. 62
%
\reference{}Hubscher, J., Paschke, A., Walter, F., 2005, IBVS 5657,
1H
%
\reference{}Hubscher, J., Paschke, A., Walter, F., 2006, IBVS 5731,
1H
%
\reference{}Hubscher, J., \& Walter, F., 2007, IBVS 5761, 1H
%
\reference{}Huth, H., 1964, Mitt. Sonnebreg 2, 112
%
\reference{}Isles, J. E., 1985a, BAA Var. Star Section Circ. No. 60
%
\reference{}Isles, J. E., 1985b, BAA Var. Star Section Circ. No. 61
%
\reference{}Isles, J. E., 1986, BAA Var. Star Section Circ. No. 63
%
\reference{}Isles, J. E., 1988, BAA Var. Star Section Circ. No. 66
%
\reference{}Kaluzny, J. and Pojmanski, G., 1984a, IBVS 2564
%
\reference{}Kaluzny, J. and Pojmanski, G., 1984b, Acta. Astron. 34,
445
%
\reference{}Kim, C.-H., Han, W., Yoon, J. H. \& Nha, I.-S., 1997, J.
Astron. Space Sci. 14, 44
%
\reference{}Kim, Chun-Hwey, Lee, Jae-Woo, Kim, Seung-Lee, Han,
Wonyong, Koch, Robert H., 2003, AJ, 125, 322K
%
\reference{}Kopal, D., 1978, Dynamics of Close binary Systems, D.
Reidel. Co.
%
\reference{}Krajci, Tom., 2005, IBVS 5592, 1K
%
\reference{}Kramer, E. H., 1948, Astr. Astrophys. 128, 84
%
\reference{}Kurpinska-Winiarska, M. and Zakrzewski, B., 1990, IBVS
3485
%
\reference{}Lanza, A. F., Rodon\`{o}, M., and Rosner, R., 1998,
MNRAS 296, 893
%
\reference{}Liu Liang, Qian Sheng-Bang, Zhu Li-Ying, He Jia-Jia,
Yuan Jin-Zhao, Dai Zhi-Bin, Liao Wen-Ping and Zhang Jia, 2008, PASJ
60, 3
%
\reference{}Liu Qingyao, Yang Yulan, Gu Shenghong, and Wang Bi,
1996, A\&As 101, 253
%
\reference{}Locher, K., 1978, BBSAG Bull. No. 39
%
\reference{}Locher, K., 1980a, BBSAG Bull. No. 46
%
\reference{}Locher, K., 1980b, BBSAG Bull. No. 51
%
\reference{}Locher, K., 1981a, BBSAG Bull. No. 52
%
\reference{}Locher, K., 1981b, BBSAG Bull. No. 53
%
\reference{}Locher, K., 1981c, BBSAG Bull. No. 56
%
\reference{}Locher, K., 1981d, BBSAG Bull. No. 57
%
\reference{}Locher, K., 1982a, BBSAG Bull. No. 58
%
\reference{}Locher, K., 1982b, BBSAG Bull. No. 62
%
\reference{}Locher, K., 1983a, BBSAG Bull. No. 64
%
\reference{}Locher, K., 1983b, BBSAG Bull. No. 65
%
\reference{}Locher, K., 1983c, BBSAG Bull. No. 69
%
\reference{}Locher, K., 1984a, BBSAG Bull. No. 70
%
\reference{}Locher, K., 1984b, BBSAG Bull. No. 71
%
\reference{}Locher, K., 1984c, BBSAG Bull. No. 74
%
\reference{}Locher, K., 1985, BBSAG Bull. No. 78
%
\reference{}Locher, K., 1986, BBSAG Bull. No. 79
%
\reference{}Locher, K., 1987a, BBSAG Bull. No. 82
%
\reference{}Locher, K., 1987b, BBSAG Bull. No. 83
%
\reference{}Locher, K., 1988a, BBSAG Bull. No. 86
%
\reference{}Locher, K., 1988b, BBSAG Bull. No. 87
%
\reference{}Locher, K., 1988c, BBSAG Bull. No. 88
%
\reference{}Locher, K., 1989a, BBSAG Bull. No. 90
%
\reference{}Locher, K., 1989b, BBSAG Bull. No. 91
%
\reference{} Lucy, L. B., Wilson, R. E., 1979, ApJ 231, 502L
%
\reference{}Lu Wenxian, 1991, AJ 102, 262L
%
%\reference{}Michaels, E. J., 1988, IBVS 3202
%
\reference{}Muyesseroglu, Z., Burol, B., and Selam, S. O., 1996,
IBVS 4380
%
\reference{}Nelson, R. H., 2001, IBVS 5040
%
\reference{}Nelson, R. H., 2007, IBVS 5760
%
\reference{}Niarchos, P. G., Hoffmann, M., Duerbeck, H. W., 1994,
A\&A 292, 494N
%
%\reference{}Niarchos, P. G., Hoffmann, H., and Duerbeck, H. K.,
%1996, A\&As 117, 105
%
\reference{}Odynskaya, O. K., 1949, Perem. Zvezdy 6, 316
%
\reference{}Ogloza, W., 1995, IBVS 4263
%
\reference{}Ogloza, W., 1997, IBVS 4534
%
\reference{}Pohl, E., Hamzaoglu, E., Gudur, N. \& Ibanoglu, C.,
1983, IBVS 2385
%
\reference{}Pohl, E., Tunca, Z., Gulmen, O. \& Evren, S., 1985, IBVS
2793
%
\reference{}Pohl, E., Akan, M. C., Sezer, C. \& Gudur, N., 1987,
IBVS 3087
%
\reference{}Pribulla, T., Vanko, M., Parimucha, S. \& Chochol, D.,
2001, IBVS 5056
%
\reference{}Pribulla, T., Vanko, M., Parimucha, S., Chochol, D.,
2002, IBVS 5341, 1P
%
%\reference{}Qian Shengbang, Liu Qingyao, and Yang Yulan, 1999, A\&A
%341, 799
%
\reference{}Qian, S.-B., 2001a, MNRAS 328, 635
%
\reference{}Qian, S.-B., 2001b, MNRAS 328, 914
%
\reference{}Qian, S.-B., 2002, A\&A 384, 908
%
\reference{}Qian, S.-B., 2003a, MNRAS 342, 1260
%
\reference{}Qian, S.-B., 2003b, A\&A 400, 649
%
\reference{}Qian, S.-B., He, J.-J., Xiang, F.-Y., Ding, X.-Y.,
Soonthornthum, B., 2005, AJ 129, 1686Q
%
\reference{}Qian, S.-B., Zhu, L.-Y., Boonruksar, S., 2006, NewA 11
503Q
%
\reference{}Qian, S.-B., Xiang, F.-Y., Zhu, L.-Y., Dai, Z.-B., He,
J.-J., Yuan, J.-Z., 2007, AJ 133, 357Q
%
\reference{}Rainger, P. P., Bell, S. A., Hilditch, R. W., 1992,
MNRAS, 254, 568R
%
\reference{}Russo, G., Sollazzo, C., Maceroni, C., Milano, L., 1982,
A\&AS 47, 211R
%
\reference{}Rucinski, S. M., 1969, A\&A 19, 245
%
\reference{}Samec, Ronald G., Su, Wen, Dewitt, Jason R., 1993, PASP
105, 1441S
%
\reference{}Selam, S. O., Gural, B., and Muyesseroglu, Z., 1999,
IBVS 4670
%
\reference{}Tout, C. A. \& Hall, D. S., 1991, MNRAS 253, 9T
%
\reference{}Wilson R. E. 1990, ApJ 356, 613
%
\reference{}Wilson R. E. 1994, PASP 106, 921
%
\reference{}Wilson R. E. \& Devinney, E. J., 1971, ApJ 166, 605
%
\reference{}Wilson R. E. \& Van Hamme, W., 2003, Computing Binary
Stars Observables, the 4th edition of the W-D programe.
%
\reference{}Voges, W., Aschenbach, B., Boller, Th., Brauninger, H.,
Briel, U., Burkert, W., Dennerl, K., Englhauser, J., Gruber, R.,
Haberl, F., and 8 coauthors, 2000, yCat 33560445B
%
\reference{}Zhukov, G. V., 1975, Astron. Tsirk. No. 888, 7







\end{references}
\end{document}